\documentclass[english,12pt]{article}
\usepackage{array}
\usepackage{graphicx}
\usepackage{amssymb}
\usepackage[english]{babel}
\usepackage{amsmath}
\usepackage{multirow}
\usepackage{prettyref}
\usepackage{babel}
\usepackage{units}
\usepackage[latin1]{inputenc}
\usepackage{amsfonts}
\usepackage{amssymb}
\usepackage{babel}
\usepackage{color}
\usepackage{cite}
\def\@fmsl@sh#1#2#3{\m@th\ooalign{$\hfil#1\mkern#2/\hfil$\crcr$#1#3$}}
 \def\eq#1\en{\begin{equation}#1\end{equation}}
\def\s[#1,#2]{[#1\stackrel{\star}{,}#2]}
\def\sx[#1,#2]{[#1\stackrel{\star_{x}}{,}#2]}

\newcommand{\nc}{\newcommand}
\nc{\beq}{\begin{equation}}
\nc{\eeq}{\end{equation}}
\nc{\beqa}{\begin{eqnarray}}
\nc{\eeqa}{\end{eqnarray}}

\def\bc{\begin{center}}
\def\ec{\end{center}}

\def\gsim{\mathrel{\mathpalette\atversim>}}

\def\bc{\begin{center}}
\def\ec{\end{center}}

\def\gsim{\mathrel{\rlap{\lower4pt\hbox{\hskip1pt$\sim$}}

    \raise1pt\hbox{$>$}}}       

\def\gsim{\mathrel{\rlap{\lower4pt\hbox{\hskip1pt$\sim$}}
    \raise1pt\hbox{$>$}}}       

\begin{document}
\makeatletter
\def\fmslash{\@ifnextchar[{\fmsl@sh}{\fmsl@sh[0mu]}}
\def\fmsl@sh[#1]#2{%
  \mathchoice
    {\@fmsl@sh\displaystyle{#1}{#2}}%
    {\@fmsl@sh\textstyle{#1}{#2}}%
    {\@fmsl@sh\scriptstyle{#1}{#2}}%
    {\@fmsl@sh\scriptscriptstyle{#1}{#2}}}
\def\@fmsl@sh#1#2#3{\m@th\ooalign{$\hfil#1\mkern#2/\hfil$\crcr$#1#3$}}
\makeatother

\thispagestyle{empty}
\begin{titlepage}
\boldmath
\begin{center}
  \Large {\bf   Three Waves for Quantum Gravity}
    \end{center}
\unboldmath
\vspace{0.2cm}
\begin{center}
{  {\large Xavier Calmet}\footnote{x.calmet@sussex.ac.uk}$^{a,b}$} {\large and}  
{  {\large Boris Latosh}\footnote{b.latosh@sussex.ac.uk}$^{a,c}$} 
 \end{center}
\begin{center}
$^a${\sl Department of Physics and Astronomy, 
University of Sussex, Brighton, BN1 9QH, United Kingdom
}\\
$^b${\sl PRISMA Cluster of Excellence and Mainz Institute for Theoretical Physics, Johannes Gutenberg University, 55099 Mainz, Germany }\\
$^c${\sl Dubna State University,
Universitetskaya str. 19, Dubna 141982, Russia}
\end{center}
\vspace{5cm}
\begin{abstract}
\noindent
Using effective field theoretical methods, we show that besides the already observed gravitational waves, quantum gravity predicts two further massive classical fields leading to two new massive waves. We set a limit on the masses of these new modes using data from the E\"ot-Wash experiment. We point out that the existence of these new states is a model independent prediction of quantum gravity. We then explain how these new classical fields could impact astrophysical processes and in particular the binary inspirals of neutron stars or black holes. We calculate the emission rate of these new states in binary inspirals astrophysical processes.
\end{abstract}  
\end{titlepage}



\newpage

Much progress has been made in recent years in quantum gravity using effective field theory methods. These methods enable one to perform quantum gravitational calculations for processes taking place at energies below the Planck mass, or some $10^{19}$ GeV while remaining agnostic about the underlying theory of quantum gravity. One could argue that the first attempts in that direction were due to Feynman who has calculated quantum amplitudes using linearized general relativity \cite{Feynman:1963ax}. Modern effective field theory techniques were introduced in the seminal works of Donoghue in the 90's \cite{Donoghue:1994dn,Donoghue:1993eb,BjerrumBohr:2002kt}. With time, it became clear that some model independent predictions could be obtained \cite{Calmet:2013hfa,Donoghue:2014yha,Calmet:2015dpa,Alexeyev:2017scq,Calmet:2008tn,Calmet:2017qqa,Calmet:2016sba,Calmet:2017rxl}. This approach is very generic and it could be the low energy theory for virtually any theory of quantum gravity such as e.g. string theory \cite{Green:1987sp,Green:1987mn}, loop quantum gravity \cite{Rovelli:1997yv}, asymptotically safe gravity \cite{Weinberg,Reuter:1996cp,Percacci:2007sz} or super-renormalizable  quantum gravity  \cite{Modesto:2011kw,Modesto:2014lga,Modesto:2015lna} just to name a few.

In this paper we point out that the low energy spectrum of quantum gravity must contain two new classical fields besides the massless classical graviton that has recently been observed in the form of gravitational waves \cite{Abbott:2016blz,Abbott:2016nmj,Abbott:2017vtc}. These new states correspond to massive objects of spin-0 and spin-2. As we will show these new states are purely classical fields that could have interesting consequences for different branches of physics, from particle physics and astrophysics to cosmology.

To identify these new fields, we calculate the leading quantum gravitational corrections to the Newtonian gravitational potential using effective field theory methods. These corrections can be shown to correspond to two new classical states that must exists besides the massless spin-2 classical graviton.  We set limits on the masses of these classical fields using data from the E\"ot-Wash pendulum experiment \cite{Hoyle:2004cw} and we then turn our attention to astrophysical and cosmological probes of quantum gravity studying quantum gravitational contributions to the inspirals of neutron stars or black holes. We demonstrate that the new massive spin-2 and spin-0 states  predicted in a model independent way by quantum gravity can modify the potential between the two astrophysical bodies and lead to testable effects. We comment on the implications of quantum gravity for inflation, dark matter and gravitational wave production in phase transition.

Although general relativity is in many regards similar to the gauge theories describing the electroweak and strong interactions, there is one basic difference which is the source of a technical difficulty with quantum gravity. The main obstacle is that the coupling constant, in the case of gravity, is a dimensional full parameter, namely Newton's constant $G_N$ while in the case of the other interactions the fundamental coupling constant is a dimensionless parameter. The fact that Newton's constant carries a dimension leads to problems with the renormalization of the theory of quantum gravity, at least at the perturbative level. While having a renormalizable theory is necessary to claim to have a fundamental theory of quantum gravity, and to perform calculations at energies above the Planck mass $M_P =1/\sqrt{G_N}\sim 10^{19}$ GeV, it is now well appreciated that using effective theory techniques leads to very interesting insights into a theory of quantum gravity \cite{Donoghue:1994dn,Donoghue:1993eb,Calmet:2013hfa,Calmet:2015dpa,Alexeyev:2017scq}. As a matter of fact, since all experiments, astrophysical or cosmological events we are aware of involve energies below the Planck mass, an effective theory of quantum gravity valid up to $M_P$ may be all that we ever need.

From a technical point of view, calculations in quantum gravity using effective theory techniques are rather simple. One integrates out the quantum fluctuations of the metric to obtain an effective action. Matter fields, depending on the problem at hand and in particular on the energy involved in the problem, can also be integrated out. One is left with an effective action given by
\begin{eqnarray}\label{action1}
S &=& \int d^4x \, \sqrt{-g} \left[ \left( \frac{1}{2}  M^2 + \xi H^\dagger H \right)  \mathcal{R}- \Lambda_C^4 + c_1 \mathcal{R}^2 + c_2 \mathcal{R}_{\mu\nu}\mathcal{R}^{\mu\nu}+ c_4   \Box \mathcal{R}  \right . \nonumber \\
&& \left . + b_1 \mathcal{R} \log \frac{\Box}{\mu^2_1}\mathcal{R} + b_2 \mathcal{R}_{\mu\nu}  \log \frac{\Box}{\mu^2_2}\mathcal{R}^{\mu\nu}  
+ b_3 \mathcal{R}_{\mu\nu\rho\sigma}  \log \frac{\Box}{\mu^2_3}\mathcal{R}^{\mu\nu\rho\sigma} 
+ \mathcal{L}_{SM}+ \mathcal{O}(M_\star^{-2})   \right],
\end{eqnarray}
where $\mathcal{R}$, $\mathcal{R}_{\mu\nu}$ and $\mathcal{R}_{\mu\nu\rho\sigma}$ are respectively the Ricci scalar, the Ricci tensor and the Riemann tensor.   The cosmological constant is denoted by $\Lambda_C$. The scales $\mu_i$ are renormalization scales which in principle could be different, we shall however take $\mu_i=\mu$. The Lagrangian $L_{SM}$ contains all of the matter we know of and $M_\star$ is the energy scale up to which we can trust the effective field theory. Note that we have written down all dimension four operators which have dimensionless coupling constants and we have thus introduced a non-minimal coupling of the Higgs doublet to curvature on top of the purely gravitational terms. The term $\Box \mathcal{R}$ is a total derivative and thus does not contribute to the equation of motions. Remarkably, the values of the parameters $b_i$ are calculable from first principles and are model independent predictions of quantum gravity, see e.g. \cite{Birrell:1982ix} and references therein. They are related to the number of fields that have been integrated out. The non-renormalizability  of the effective action is reflected in the fact that we cannot predict the coefficients $c_i$ which, in this framework, have to be measured in experiments or observations. There will be new $c_i$ appearing at every order in the curvature expansion performed when deriving this effective action and we thus would have to measure an infinite number of parameters. Despite this fact, the effective theory leads to falsifiable predictions as the coefficients $b_i$ of  non-local operators are, as explained previously, calculable.

The effective action contains three classical fields: the well known massless spin-2 field (the classical graviton) $h^{\mu\nu}$, a massive spin-2 classical field $k^{\mu\nu}$  and a massive classical spin-0 field $\sigma$ on top of the mater fields contained in $\mathcal{L}_{SM}$. This can be see explicitly by sandwiching the Green's function of the metric  in the linearized effective action between two classical sources $T^{(i)\mu\nu}$ \cite{Calmet:2017rxl}
\begin{eqnarray} \label{fc}
 256 \pi^2 G_N^2 &&  \left [ \frac{T^{(1)}_{\mu\nu} T^{(2)\mu\nu} -    \frac{1}{2} T^{(1)\mu}_{\ \ \mu} T^{(2)\nu}_{\ \ \nu}}{k^2}
-\frac{T^{(1)}_{\mu\nu} T^{(2)\mu\nu} - \frac{1}{3} T^{(1)\mu}_{\ \ \mu} T^{(2)\nu}_{\ \ \nu}}{k^2-\frac{2}{ \kappa^2 \left(c_2+\left (b_2+ 4 b_3 \right ) \log \left( \frac{-k^2}{\mu^2} \right )\right)}} \right .
\\ \nonumber && \left .
+ \frac{T^{(1)\mu}_{\ \ \mu} T^{(2)\nu}_{\ \ \nu}}{k^2-\frac{1}{\kappa^2 \left (3 c_1 + c_2+\left(3 b_1+ b_2+ b_3 \right)\log\left(\frac{-k^2}{\mu^2}\right) \right )}}
 \right ],
\end{eqnarray}
where $\kappa^2=32 \pi G$. A careful reader will have noticed the minus sign in front of the massive spin-2 mode. This is the well known ghost due to the the term $R_{\mu\nu}R^{\mu\nu}$. However, the corresponding state $k^{\mu\nu}$ is purely classical and it does not lead to any obvious pathology.  This is simply a repulsive classical force. We will show that the emission of this massive spin-2 wave leads to the production of waves with positive energy. This state simply effectively couples with a negative coupling constant $M_P$  to matter. It is crucial to appreciate that this mode is purely classical and should not be quantized as it is obtained by integrating out the quantum fluctuations of the graviton from the original action.

Using Eq.(\ref{fc}), it is straightforward to calculate the leading second order in curvature quantum gravitational corrections to Newton's potential  of a point mass $m$. We find: 
\begin{eqnarray}
\Phi(r) = -\frac{Gm}{r} \left( 1+\frac{1}{3}e^{-Re(m_0) r}-\frac{4}{3}e^{-Re(m_2) r} \right) 
\end{eqnarray}
where the masses are given by
\begin{align}
m_2^2=\frac{2}{ (b_2+ 4 b_3) \kappa^2 W\left(-\frac{2 \exp\frac{c_2}{(b_2+ 4 b_3)}}{ (b_2+ 4 b_3) \kappa^2 \mu^2}\right)},
\end{align}
\begin{align}
m_0^2=\frac{1}{ (3 b_1+b_2+ b_3) \kappa^2 W\left(-\frac{ \exp\frac{3 c_1+c_2}{(3 b_1+b_2+ b_3)}}{ (3 b_1+b_2+ b_3) \kappa^2 \mu^2}\right)},
\end{align}
and where $W(x)$ is the Lambert function. This effective Newtonian potential is a generalization of Stelle's classical result \cite{Stelle:1977ry}, it includes the non-local operators as well as the local ones and thus contains the leading quantum gravitational corrections at second order in curvature. 

Note that our result is compatible with the results obtained in \cite{Donoghue:1993eb,BjerrumBohr:2002kt,Iwasaki:1971vb}, we simply focus on a different limit where the coefficients of $R^2$ and $R_{\mu\nu} R^{\mu\nu}$ are not necessarily tiny. It is easy to show that the effective action leads to higher order corrections in $G_N$  to the Newtonian potential energy of two large non-relativistic masses $m_1$ and $m_2$. The quantum corrected Newtonian potential is given by
\begin{eqnarray}
U(r)=- G_N \frac{m_1 m_2}{r}
- 3  G_N^2 \frac{m_1 m_2 (m_1+m_2)}{r^2}
- \frac{m_1 m_2}{\pi r^3} G^2_N \left ( \frac{N_s}{42} + \frac{N_f}{7} + \frac{2 N_V}{7} + \frac{41}{10}\right).
\end{eqnarray}
This extends the result presented in \cite{Donoghue:1993eb,BjerrumBohr:2002kt,Iwasaki:1971vb} to include the numbers $N_i$ respectively of real scalar fields, Dirac fermions and vector fields present in the model. The number of matter fields $N_i$ are related to the  $b_i$ which are the Wilson coefficients appearing in the effective action by the relations  $N_i= b_{2,i}+ 4 b_{3,i}$. Here we took the same limit as in \cite{Donoghue:1993eb,BjerrumBohr:2002kt,Iwasaki:1971vb} assuming that the $c_i$ are very small. The corresponding terms lead to delta functions which do not contribute to the potential energy. As emphasized in \cite{BjerrumBohr:2002kt}, the second term in the potential represents the leading relativistic correction and it is not a quantum correction. Note that these corrections are appearing at order $G_N^2$ and are thus subleading in comparison to the contributions of the new waves appearing in $\Phi(r)$ on which we will thus focus.

The masses of the new modes correspond to pairs of complex poles in the green's functions of the massive spin-2 $k^{\mu\nu}$ and spin-0 $\sigma$ states. In general, the masses are complex depending on the values of the parameters $c_i$, $b_i$ and $\mu$, in other words they contain a width. The imaginary contributions, however, vanish when adding up the contributions of these states to the Newtonian potential. It is straightforward to show that Stelle's classical result is recovered in the limit of $b_i=0$. 

It is easy to work out the coupling of $k^{\mu\nu}$ and $\sigma$ to matter. We find
\begin{align}
S=\int d^4 x \left[\left (- \frac{1}{2} h_{\mu\nu} \Box h^{\mu\nu}
 +\frac{1}{2} h_{\mu}^{\ \mu} \Box h_{\nu}^{\ \nu}  -h^{\mu\nu} \partial_\mu \partial_\nu h_{\alpha}^{\ \alpha}+ h^{\mu\nu} \partial_\rho \partial_\nu h^{\rho}_{\ \mu}\right) \right. \\ 
\left. \nonumber + \left ( -\frac{1}{2} k_{\mu\nu} \Box k^{\mu\nu}
 +\frac{1}{2} k_{\mu}^{\ \mu} \Box k_{\nu}^{\ \nu}  -k^{\mu\nu} \partial_\mu \partial_\nu k_{\alpha}^{\ \alpha}+ k^{\mu\nu} \partial_\rho \partial_\nu k^{\rho}_{\ \mu}
 \right.  \right.  \\ \left. \left. \nonumber
 -\frac{m_2^2}{2} \left (k_{\mu\nu}k^{\mu\nu} - k_{\alpha}^{\ \alpha} k_{\beta}^{\ \beta} \right )
 \right)  \right.
  \\
  \nonumber
 \left. + \frac{1}{2} \partial_\mu \sigma  \partial^\mu \sigma
  - \frac{m_0^2}{2} \sigma^2 - \sqrt{8 \pi G_N} (h_{\mu\nu}-k_{\mu\nu}+\frac{1}{\sqrt{3}} \sigma \eta_{\mu\nu})T^{\mu\nu} 
  \right ].
\end{align}
This result shows that quantum gravity, whatever the underlying ultra-violet theory might be, has at least three classical degrees of freedom in its low energy spectrum. The massless mode has recently been directly observed in the form of gravitational waves. While there was little doubt about their existence since the discovery of the first binary pulsar in 1974, the direct observation by the LIGO and Virgo collaborations \cite{Abbott:2016blz,Abbott:2016nmj,Abbott:2017vtc}  erased any possible remaining doubt. While the massless mode affects the distance between two points, and thus the geometry, the massive modes are of the 5th force type and they do not affect the geometry of space-time. A 5th force will not change the proper distance between the mirrors of an interferometer such as those of LIGO or Virgo, but it could still lead to measurable displacement of the mirrors if the wavelength is shorter than the distance between the mirrors on one arm of an interferometer.

 We find that the strength of the interaction between the new massive modes and matter is fixed by the gravitational coupling constant. It is crucial to appreciate that the fields $h^{\mu\nu}$, $k^{\mu\nu}$ and $\sigma$ are purely classical degrees of freedom. This is why the overall negative sign of the kinetic term of $k^{\mu\nu}$ is not an issue, it simply implies that this field couples with a negative Planck mass to matter. We shall demonstrate that the corresponding massive spin-2 wave produced in binary inspiral does not violate energy conservation. Note that while $k^{\mu\nu}$ couples universally to matter, $\sigma$ does not couple to massless vector fields \cite{Hindawi:1995an,Calmet:2017voc}. 

The fact that these fields are purely classical has some interesting consequences if one tries to interpret the massive modes as dark matter candidates or the inflaton in the case of the scalar field. If the massive modes constitute all of dark matter, dark matter would an emergent phenomenon. In that sense dark matter would be fundamentally different from regular matter. 
The same remark applies to inflation if the scalar field encompassed in the curvature squared term is responsible for the early universe exponential expansion.

We now turn our attention to the experimental bounds on the masses of the two heavy states. Newton's potential with its quantum gravitational corrections can be probed with sub-millimeter tests of the gravitational inverse-square law \cite{Hoyle:2004cw}. In the absence of accidental fine cancellations between both Yukawa terms,  the current bounds imply $m_0$ , $m_2 > (0.03 \ \mbox{cm})^{-1}=6.6 \times 10^{-13}$GeV. Note that the E\"ot-Wash  experiment performed by Hoyle et al. \cite{Hoyle:2004cw} is probing separations between 10.77 mm and 137  $\mu$m, a cancelation between the two Yukawa terms on this range of scales seems impossible without modifying general relativity with new physics to implement a screening mechanism.

The bound on the quantum gravitational corrections to Newton's potential imply that quantum gravity could only impact the final moments of the inspiraling  of binary of two neutron stars or of two black holes. Their effect will only become relevant at distances shorter than $0.03 \ \mbox{cm}$.  There are two possible effects. When the two astrophysical bodies are close enough, Newton's law could be affected by the propagations of the new massive modes and the new massive modes could be produced in the form of new massive waves. 

The quantum gravitational correction to the orbital frequency  of a inspiraling binary system is given by
\begin{eqnarray}
\omega^2= \frac{Gm}{r^3} \left( 1+\frac{1}{3}e^{-Re(m_0) r}-\frac{4}{3}e^{-Re(m_2) r} \right) 
\end{eqnarray}
where $m=m_1+m_2$ is the total mass of the binary system. The total energy of the system is given by
\begin{eqnarray}
E=- \frac{Gm \mu}{2 r} \left( 1+\frac{1}{3}e^{-Re(m_0) r}-\frac{4}{3}e^{-Re(m_2) r} \right) 
\end{eqnarray}
where $\mu=m_1 m_2/m$ is the reduced mass of the system. The quantum gravitationally corrected waveform can be deduced from the energy-conservation equation $\dot E=-P_{GW}$ where $P_{GW}$ is the power of the quadrupole radiation of the gravitational waves corresponding to the massless spin-2 mode:
\begin{eqnarray}
P_{GW}=\frac{32 G_N \mu^2 \omega^6 r^4}{5 c^2}
\end{eqnarray}
which can be solve for $r(t)$ from which $\omega(t)$ can be calculated. The quantum corrected chirp signal which has frequency $f_{GW}$ and amplitude $A_{GW}$ can then be obtained in a straightforward manner:
\begin{eqnarray}
f_{GW}(t)&=&\frac{\omega(t)}{\pi} \\
A_{GW}(t)&=&\frac{1}{d_L} \frac{2 G_N}{c^4} 2 \mu \omega(t) r^2(t),
\end{eqnarray}
where $d_L$ is the luminosity distance of the source.

While it is easy to calculate $f_{GW}$ and $A_{GW}$ explicitly, it is clear that the quantum gravitational corrections to the emission of gravitational waves can only become relevant when the two objects are closer than $0.03 \ \mbox{cm}$ given the bound derived on the mass of the massive spin-2 object using data from the E\"ot-Wash experiment \footnote{The effects of the $1/r^2$  and $1/r^3$ terms discussed above, which are corrections to the propagation of the massless mode will be considered elsewhere\cite{CEM}.}. This distance is well within the Schwarzschild radius of any astrophysical black hole and clearly tools from numerical relativity need to be employed to obtain a reliable computation. Note that for black holes the mass is concentrated at their center and very close to the singularity. While the horizons will have started to merge, the two singularities could be within a reasonable distance of each other. In that sense our approximation may not be so rough. In any case it is clear that incorporating our quantum gravitational effect in numerical relativity calculations \cite{Bishop:2016lgv} represents a real technical challenge as the interior of black holes is usually excised to avoid having to discuss the singularities. However, the new states can only be relevant when the distance between the two black hole singularities become of the order of the inverse of the mass of the massive spin-2 object.

Besides the usual massless gravitational waves, there are two new kind of radiations, namely the massive spin-0 and spin-2 could in principle be produced in energetic astrophysical or cosmological events. However, in the case of a binary system, because the center of mass of the system is conserved, the spin-0 wave cannot be produced. On the other hand, the massive spin-2 could be emitted in the last moment of a merger when the two inspiraling objects are closer than the inverse of the mass of the massive spin-2 field. A lengthy calculation leads to a remarkable result. The energy $E$ carried away by the massive spin-2 mode from a binary system per frequency is identical to that of massless spin-2 mode:
\begin{eqnarray}
\frac{d E_{massive}}{d\omega}=\frac{G_N}{45} \omega^6 \langle  Q_{ij}  Q^{ij} \rangle \theta(\omega-m_2)
\end{eqnarray}
 up to a Heaviside step function which prevents the emission of massive waves when the energy of the system is below the mass threshold. Note that as usual $Q_{ij}$ is the quadrupole moment of the binary system. The total wave emission by a binary system is thus given by
\begin{eqnarray}
\frac{d E}{d\omega}=\frac{d E_{massless}}{d\omega}+\frac{d E_{massive}}{d\omega},
\end{eqnarray}
where the first term on the righthand side is the usual general relativity result for massless gravitational waves. Once the massive channel becomes available, half of the energy is damped into the massive mode.

The massive spin-2 wave will only be produced when the two black holes are close enough from another. If we denote the distance between the black holes of masses $m_A$ and $m_B$ by $d$, we obtain the frequency of the inspiral $\omega$:
\begin{eqnarray}
  \omega^2 = \cfrac{G_N(m_A+m_B)}{d^3}.
\end{eqnarray}
To estimate how close the two black holes have to be to generate enough energy to produce a massive wave compatible with the E\"ot-Wash bound, we set $\omega=(0.03 \ \mbox{cm})^{-1}$ and use the masses of the first merger observed by the LIGO collaboration  $m_A =36 M_\odot$ $m_B=29 M_\odot$ (where $M_\odot$ is the mass of the sun). We find that for a wave of mass $(0.03 \ \mbox{cm})^{-1}$ to be produced the two black holes would have to be at $16 \ \mbox{cm}$ from another. Clearly this is again well within the horizon of any astrophysical black holes and a reliable simulation will require a challenging numerical investigation. In any case, our results demonstrate that massive spin-2 waves can be produced in the merger of astrophysical objects such as black holes and this effect must be taken into account in future numerical studies. Clearly the massive modes will only be produced in the final stage of the inspiral process at the time of the merger and ringdown. This represents a unique opportunity to probe quantum gravity with astrophysical events in a fully non-speculative manner.

Let us emphasize at this stage that we have considered binary systems in the Newtonian regime. Our main motivation was to demonstrated that first principle quantum gravitational calculations are possible.  It is, however, clear that the leading order correction that we have considered here cannot be trusted in the insprial process when two astrophysical objects reach very short distances and higher order post-Newtonian corrections or, more likely, a full numerical general relativity becomes necessary. Let us also stress that we have considered the most optimistic case scenario, still compatible with the E\"ot-Wash experiment, by studying masses for the new fields of the order of $(0.03 \ \mbox{cm})^{-1}$. However, the masses of these new fields could be anywhere between $(0.03 \ \mbox{cm})^{-1}$  and the inverse Planck length or some  $(1.6 \times 10^{-35} \ \mbox{m})^{-1}$. If numerical studies managed to consider distances equal or shorter to $(0.03 \ \mbox{cm})^{-1}$, then gravitational signals from binary system would enable one to probe quantum gravity more accurately than the E\"ot-Wash experiment.

As mentioned previously, such short distances are well within the Schwarzschild radius of any astrophysical body. This implies that mergers of neutron stars are unlikely to enable one a probe of quantum gravity. On the other hand, depending on how we think of black holes, binary systems of such objects might enable one to probe very short distance. Astrophysical black holes are the end product of the gravitational collapse of matter such as e.g. stars. Under such a collapse, matter falls towards the singularity but we expect quantum physics to smear out the singularity. In that sense, one expects the gravitational collapse of matter to lead to a very dense ring of matter at the center of the black hole. We can thus think of a black hole as an extremely dense object with matter concentrated within a Planck length of the center of the black hole. The horizon itself is not a physical object, a falling observer never notices that he passes through the horizon. It is simply a reaction of space-time to the presence of the very dense core of the black hole. While physical phenomena taking place within the horizon cannot be observed directly by an external observer, the horizon would react to a change in the matter distribution inside such an horizon. We can thus think of a black hole merger as the merger of two extremely dense astrophysical bodies. When the two dense cores get close enough, a common horizon forms, this common horizon will keep on evolving as a the two cores continue to move towards each other inside the common horizon. This is not the standard picture which usually solely focusses on the dynamics of the horizon (indeed numerical studies usually excise space-time inside the horizon), but it must be equivalent. On the other hand, thinking of black holes as extremely dense core objects with an horizon that is a response of space-time to this dense center would enable one to study extremely short distance physics, potentially up to the Planck length. This is not doable in  standard numerical studies which artificially remove the inside of black holes, purely for technical reasons. The feasibility of this alternative approach will be investigated elsewhere.

While we discussed the production of the massive waves in the context of astrophysical processes, it is also possible to envisage the production of these new quantum gravitational massive classical modes during first order phase transitions if such phases took place early on in the cosmological evolution of our universe. Clearly, the occurrence of a first order phase transition in the early universe is a speculative topic as there is no such phase transition within the electroweak standard model. Our work represents an additional complication for the study of early universe phase transitions as beyond the massless gravitational waves, the new massive modes could be produced. Indeed, the collision of bubbles and damping of plasma inhomogeneities could have generated a stochastic background of massive gravitational waves beyond the massless ones that are expected. This implies that some of the energy of these processes could be lost in massive modes. This fact has been overlooked so far when doing simulations for LISA \cite{Caprini:2015zlo}. 

Tests of quantum gravity often focus on exotic possibilities \cite{Hossenfelder:2010zj}  such as the presence of Lorentz violation effects \cite{Kostelecky:2003fs} or other kinds of symmetry breaking. In the case of gravitational waves, different extensions of general relativity \cite{DeLaurentis:2016jfs,Bogdanos:2009tn,Jones:2016zqw,Andriot:2017oaz} have been considered. In this paper, we have shown that there are model independent predictions of quantum gravity which can be searched for in experiments or in observations. The main prediction is the existence of two new classical states namely a massive spin-2  and massive spin-0 classical fields. The phenomenology of these fields is clear, their interactions with matter is fixed by the underlying theory of quantum gravity. The only unknown parameters are their masses.  It is thus essential to study these states and hopefully to discover them in an experiment or observation. This program is extremely conservative as any theory of quantum gravity must at least contain these two new states beyond massless gravitational waves. While we cannot calculate their masses from first principles, we have shown that there are bounds on the masses of these new classical fields. This approach to quantum gravity opens up new directions to understand dark matter and inflation which could be emergent, i.e., purely classical, phenomena.

{\it Acknowledgments:}
The work of XC is supported in part  by the Science and Technology Facilities Council (grant number  ST/P000819/1). XC is very grateful to MITP for their generous hospitality during the academic year 2017/2018. 


\bigskip{}

\end{document}